\documentclass[10pt,conference]{IEEEtran}

  \usepackage{fancyhdr}
  \usepackage[left=0.65in,right=0.65in,top=0.67in,bottom=0.98in]{geometry}
  \usepackage{cite}
  \usepackage{url}
  \usepackage{amsmath}
  \usepackage{tabularx}
  \usepackage{subfigure}
  \usepackage{amsmath}
  \usepackage{breqn}
  \usepackage{mathtools}
  \usepackage{cuted}
  
\ifCLASSINFOpdf
  \usepackage{graphicx}
\else
\fi
\graphicspath{ {./figures/} }
\usepackage{subfigure}
\usepackage[font = footnotesize]{caption}
\usepackage{ifpdf}
\usepackage{graphicx}
\usepackage{amsmath}
\usepackage{pslatex}
\usepackage{xcolor}
\usepackage{textcomp}
\usepackage{placeins}
\usepackage{eqparbox}
\usepackage{url}
\usepackage{stfloats}
\usepackage{multicol}
\usepackage{lipsum}
\usepackage{float}
\usepackage{algorithmicx}
\usepackage{algpseudocode}
\usepackage[linesnumbered,ruled,vlined]{algorithm2e}
\usepackage{amsthm}
\usepackage{comment}
\usepackage{cite}
\usepackage{enumerate}
\usepackage{array}
\usepackage{balance}
\usepackage{color}
\usepackage{amssymb,amsmath}
\usepackage{xcolor}

\newcommand\blfootnote[1]{%
  \begingroup
  \renewcommand\thefootnote{}\footnote{#1}%
  \addtocounter{footnote}{-1}%
  \endgroup
}

\DeclareFontFamily{U}{matha}{\hyphenchar\font45}
\DeclareFontShape{U}{matha}{m}{n}{
      <5> <6> <7> <8> <9> <10> gen * matha
      <10.95> matha10 <12> <14.4> <17.28> <20.74> <24.88> matha12
      }{}
\DeclareSymbolFont{matha}{U}{matha}{m}{n}

\DeclareMathSymbol{\Lt}{3}{matha}{"CE}
\DeclareMathSymbol{\Gt}{3}{matha}{"CF}

\begin{document}

\title{On the Spatial Consistency of Sub-Terahertz Channel Characteristics for Beyond-6G Systems}

\author{
\IEEEauthorblockN{
Hossein Amininasab$^{1}$, Huda Farooqui$^{1}$, Dmitri Moltchanov$^{1}$, Sergey Andreev$^{1}$, \\ Michele Polese$^{2}$, Mikko Valkama$^{1}$, Josep M. Jornet$^{2}$}

\IEEEauthorblockA{
$^{1}$Unit of Electrical Engineering, Tampere Wireless Research Center, Tampere University, Finland\\
$^{2}$Department of Electrical and Computer Engineering, Northeastern University, Boston, MA, USA\\Contact author's e-mail: hossein.amininasab@tuni.fi
}\vspace{-0mm}}
\maketitle

\blfootnote{\copyright~2026 IEEE. Personal use of this material is permitted. For all other uses, permission must be obtained from IEEE. This is the author's accepted manuscript of a paper accepted for publication in the proceedings of IEEE VTC-Spring 2026.}
\begin{abstract}
Ray tracing is a versatile approach for precise sub-terahertz (sub-THz, 100--300~GHz) channel modeling when designing new mechanisms for beyond-6G cellular systems. Theoretically, wireless channels may exhibit variations over wavelength distances. In the sub-THz band, close-to-millimeter wavelengths thus require extremely large computational efforts for ray-tracing modeling. However, in practice, channel characteristics may remain quantitatively similar over much larger distances, which can drastically decrease computational efforts. The aim of this study is to experimentally characterize the degree of spatial consistency in sub-THz channel characteristics. To this end, we performed a large-scale measurement campaign in the 140--150~GHz frequency band in an indoor-hall (InH) environment and characterized the channel at separation distances from 2.5~mm up to 1~m. Our results show that channel characteristics including delay spread, angular delay spread, and K-factor change only slightly over multiple tens of centimeter distances. This implies that, in the considered InH environment, the mesh grid can be in the range of 10--50 wavelengths (at 145~GHz) along stable line-of-sight (LoS) directions, while a finer resolution is needed in regions not dominated by LoS. For coarser grids, advanced interpolation is required to capture rapidly varying scattered components. 
\end{abstract}


\section{Introduction}\label{sect:intro}

The utilization of high-frequency bands, such as the sub-THz (100--300~GHz), in beyond-6G cellular systems is vital for fundamental improvements in access rates at the air interface \cite{itur6G}. These bands, providing hundreds of megahertz of contiguous bandwidth at the air interface, can potentially enable future applications characterized by extreme rate demands \cite{moltchanov2022tutorial}.

Sub-THz frequency band shares a number of propagation properties with millimeter-wave band (mm-Wave, 30--100~GHz) including diffuse diffraction, sensitivity to materials in urban deployments, blockage, etc. \cite{serghiou2022terahertz}. However, further decrease in the wavelength makes sub-THz wave propagation even more sensitive to environmental specifics such as the type of materials and geometrical properties of surrounding micro-scale objects. In the context of beyond-6G communication systems, the wavelength also affects the size of antenna elements, requiring larger arrays at base stations (BS), resulting in narrower beamwidths and making the system more sensitive to beam misalignment events \cite{humayra2025distinguishing,moltchanov20256g}.

To take decisive steps toward the development of beyond-6G systems, fine-grained propagation models are required. Conventional propagation models, such as urban-micro/macro (UMi/UMa) and indoor-hall (InH), predict the average receive signal strength at a given distance from the transmitter \cite{3gpp_los}. Although sub-THz counterparts of such models have recently been proposed \cite{han2022terahertz}, these models do not capture time-related characteristics of wireless channel and thus fail to provide detailed characterization of the propagation environment required to design physical-layer mechanisms. Three-dimensional (3D) cluster channel models inherently account for multipath propagation and random delays representing the received signal strength as a random variable, see, e.g. \cite{cheng2020thz}. However, these models require complex procedures to account for spatial consistency, such as ensuring that there is a dependency between the received signal strength observed at adjacent distances \cite{gapeyenko2019spatially}. 

Ray tracing is nowadays a de-facto approach for precise modeling of mmWave/sub-THz communications systems \cite{zhang2024deterministic}. Theoretically, channel response is expected to change every wavelength distance thus requiring extreme computational efforts to characterize realistic cellular deployments \cite{yi2022ray}. However, in practical conditions, the channel characteristics may remain approximately the same over much larger distances that may significantly simplify computational efforts or facilitate efficient interpolation techniques. Thus, it is critical to identify the distance over which channel characteristics remain approximately the same.

To the best of our knowledge, this paper presents the first experimental study on the spatial consistency of sub-THz channel characteristics at distances smaller than 1 m, conducted through a comprehensive and large-scale measurement campaign in the open scientific literature. Measurements were carried out in the 140--150~GHz frequency band within an InH environment, investigating variations in channel properties over separation distances ranging from 2.5~mm to 1~m to assess spatial consistency across multiple scales. The analysis utilizes spatial autocorrelation of key channel characteristics, including the power delay profile (PDP), root-mean-square (RMS) delay spread (DS), RMS angular spreads (AS), and K-factor. These findings provide insights for parameterizing sub-THz ray tracers for indoor propagation modeling, quantifying modeling errors, and developing channel prediction algorithms.

The main contributions of our study are:
\begin{itemize}
    \item{We present measurements and an analysis of sub-THz channel characteristics in the 140--150~GHz frequency band to assess spatial consistency in InH environments.}
    \item{We provide empirical observations showing that: (i) channel metrics such as delay spread, angular spread, and K-factor remain stable over small-scale displacements ($\leq 10–50 \lambda$), indicating strong local spatial consistency; and (ii) spatial de-correlation at larger separations primarily arises from variations in non-line-of-sight (NLoS) multipath components (MPCs), highlighting the need for adaptive meshing in ray-based modeling and for robust beam-tracking strategies in future InH sub-THz systems.}
\end{itemize}

The paper is organized as follows. We begin with related work in Section \ref{eqn:related}. The measurement setup, experiments, and data processing are described in Section \ref{sect:meas}. Results and discussion are provided in Section \ref{sect:results}. Conclusions are drawn in the last section.

\section{Related Work}\label{eqn:related}

The wave propagation at sub-THz band already received significant attention from the research community. The existing efforts in \cite{mahmood_2024_analysis, xing_2021} have focused on large-scale propagation including path loss, delay spread, and material interactions in indoor and outdoor environments. These studies report that, in the presence of LoS transmission, the Path Loss Exponent (PLE) is in the range of 1.8--2.2 depending on environmental conditions. However, in a NLoS environment, PLE could increase up to 3.6, due to diffraction loss, penetration through materials, and blockage \cite{xing_2021,kyosti_2021}. In addition, an extra attenuation of 10--30~dB is commonly observed due to beam misalignment or structural obstructions. 

Although there has been substantial effort on large-scale propagation, only few authors studied spatial or temporal consistency \cite{ma_2024,han_2022_THz}. At lower mm-Wave bands ($\leq 100$~GHz), standardized models such as 3GPP TR 38.901 and NYUSIM \cite{ju_2018} define reference correlation distance of approximately 10--15~m for shadow fading, delay spread, and angular spreads. These values are obtained from extensive measurement campaigns and serve as a reference for evaluating channel variations caused by displacement. 

At the sub-THz bands, the correlation distance is expected to decrease as displacement-induced variations increase, particularly in environments where directionality is significant. Ray-based deterministic channel modeling studies have shown that narrow-beam antennas are more likely to increase PLE, whereas wide-beam antennas reduce PLE at the expense of a longer delay spread \cite{gougeon_2019}. These observations indicate that the antenna beamwidth strongly affects channel statistics that are closely related to spatial consistency. Precise metrology in the 200--300~GHz range demonstrates that even submillimeter displacements can lead to measurable channel variations. Spatial correlation may change drastically due to environmental factors such as human body blockage, furniture, plants, and weather conditions \cite{ma_2024}, while controlled studies show that path loss, delay spread, and cluster statistics often exhibit more gradual evolution with displacement.

Recent research provides evidence of these trends. In \cite{ju_2021_140}, the authors conducted measurements at 142~GHz within an urban microcell, capturing key large- and small-scale wireless channel properties along a 102~m rectangular path, with Rx points spaced every 3~m. Each parameter’s spatial autocorrelation was modeled using an exponentially decaying sinusoid, revealing that shadow fading de-correlates rapidly, with a correlation distance of only 3.8~m, significantly shorter than the 10--13~m \cite{etsi_2020_5g} typically observed at below 100~GHz. The delay spread and the angular spread exhibit larger correlation distances, approximately 11.8~m and 12~m, respectively, similar to those seen at lower frequencies \cite{ju_2021_140}. 

Complementing these efforts, a large-scale indoor measurements campaign of 90 transmitter (Tx)-receiver (Rx) pairs \cite{chen_2023} qualitatively characterized temporal and spatial consistency, in hallway and NLoS cases, and reported per-scenario PLEs, delay/angle spreads, and cluster statistics across displacement ranges of 2--30~m. The results verify that dominant MPCs persist and angle-of-arrival (AoA) and time-of-arrival (ToA) shift smoothly along displacement tracks, supporting the feasibility of beam tracking at 200+~GHz. Indoors, several authors extended statistical models using 140~GHz datasets and developed spatially-aware simulators (e.g., NYUSIM extensions) \cite{ju_2020_3d}. These works provide rich measurement databases and cluster/delay/angular spread statistics, forming a practical basis for spatial consistency modeling at frequencies above $100$~GHz. 

Further studies reinforced the sensitivity of sub-THz channels to displacement. At 73~GHz, fading varies smoothly over 0.35~m \cite{ju_2018}. Measurements at 215--225~GHz showed delay spreads of 0.16--0.83~ns, azimuth spreads of 14.65°--37.80°, and 2.57--4.14 clusters depending on displacement \cite{wang_2024}, while at 299--301~GHz in large indoor halls, delay spreads increased sharply with distance, from $\approx 7.5$~ns at 3~m up to $\approx 35.6$~ns at 15.6~m, accompanied by angular spreads exceeding 40° and highly non-stationary multipath behavior \cite{lyu_2023}. Narrow-beam antennas reduce delay spread but raise PLE, while wide beams broaden multipath variation \cite{sen2024impact}. High-precision chamber experiments at 200--300~GHz further show that even sub-millimeter shifts cause measurable delay drift \cite{al2025}. 
Summarizing the related studies, we emphasize that little is known about the spatial consistency of sub-THz channel characteristics at distances below 1 m, which are particularly relevant for the parameterization of ray-tracing simulations. In our paper, we focus on such smaller separation distances between measurement points. We also consider InH propagation environment that is subject to complex multipath propagation.

\section{Measurement Campaign}\label{sect:meas}

In this section, we describe the measurement campaign including the measurement setup using the Northeastern University (NU) channel sounder, the experiments, and the measurement data evaluation part.

\subsection{Measurement Setup} 

\begin{figure}[t!]
    \centering
    \includegraphics[width=1\linewidth]{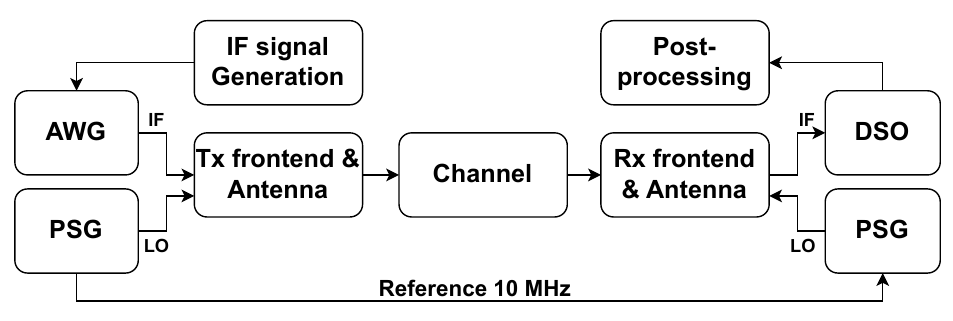}
    \caption{Block diagram of the NU channel sounder and the equipment in use for the channel sounding measurements.}
    \label{fig:BlockDiagram_ChannelSouding}
    \vspace{-3mm}
\end{figure}

Fig.~\ref{fig:BlockDiagram_ChannelSouding} illustrates block diagram of the NU channel sounder, a custom spread-spectrum sliding-correlator channel sounder developed at Northeastern University~\cite{sen2022terahertz}. For PDP measurements, WR-6.5 VDI horn antennas were employed at both Tx and Rx to radiate and capture the pseudo-random sequences with sharp autocorrelation properties over a 10~GHz bandwidth. 

The sounding sequences were generated in MATLAB, uploaded to an Arbitrary Waveform Generator (AWG), and fed into the intermediate frequency (IF) port of the VDI front-end. A Keysight performance signal generator (PSG) provided local oscillator (LO) to upconvert the signal to 140~GHz, which was radiated through a horn antenna (21~dBi gain, 13$^\circ$ full 3-dB beamwidth).

\captionsetup[figure]{labelfont={default},labelformat={default},labelsep=period,name={Fig.  }}

At the Rx, the signal was captured by an identical VDI horn antenna connected to a VDI front-end, which was synchronized with a second PSG. LO synchronization between PSGs was achieved using a 10~MHz reference cable. The downconverted IF signal was sampled at 128~GSa/s and recorded by a digital storage oscilloscope (DSO) for further processing. Considering the DSO sampling frequency, the sampling interval is $T_s= 7.8125$~ps. The fundamental delay resolution of the channel sounding is bandwidth-limited to $\Delta_\tau=1/B=200$~ps for a bandwidth of $B=10$~GHz.

The Rx front-end and horn antenna were mounted on a motorized rotary table, controlled via MATLAB through a serial interface, allowing systematic sweeps of the Rx orientation in both azimuth and elevation.

\subsection{Description of Experiments}\label{subsec:desexp}
 
\begin{figure}[t!]
    \centering
    \includegraphics[width=1.0\linewidth]{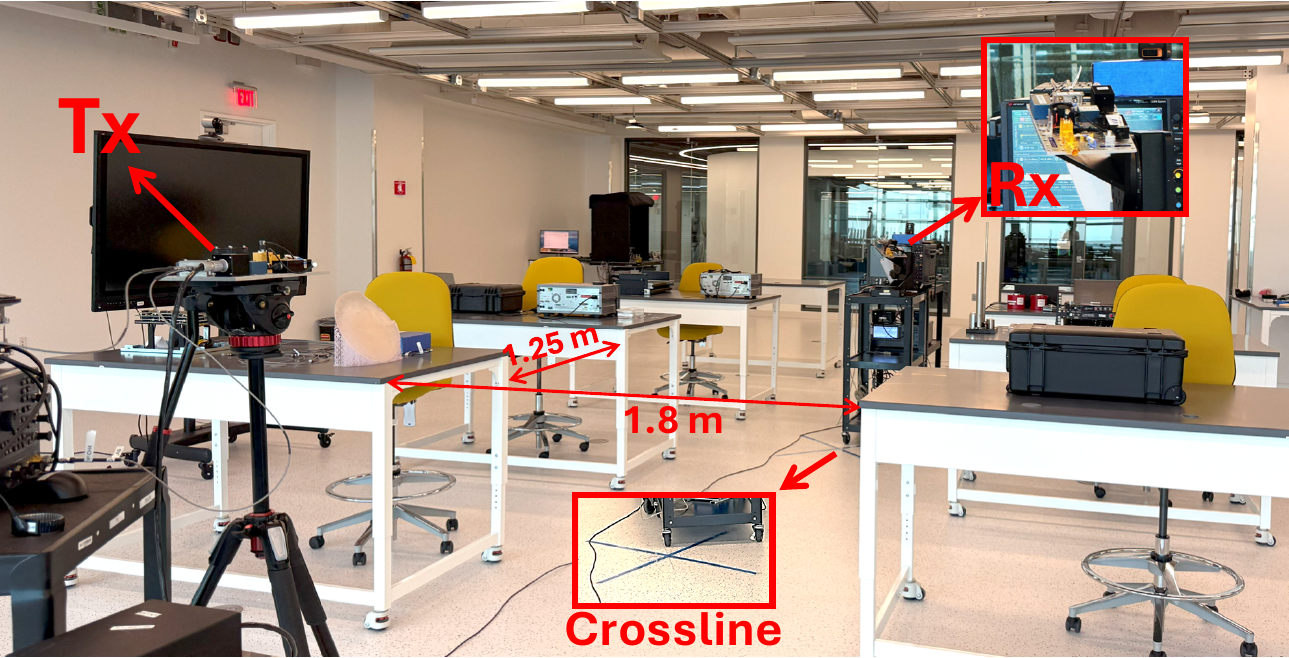}
    \caption{Channel sounding measurement environment (UNLab, EXP Building, 7th floor) resembling a classroom, showing the cross-shaped trajectory, distances between tables, and the NU channel sounder setup.}
    \label{fig:Measurement_ChannelSounding}
    \vspace{-3mm}
\end{figure}

\begin{figure}[b!]
    \vspace{-3mm}
    \centering
    \includegraphics[width=0.8\linewidth]{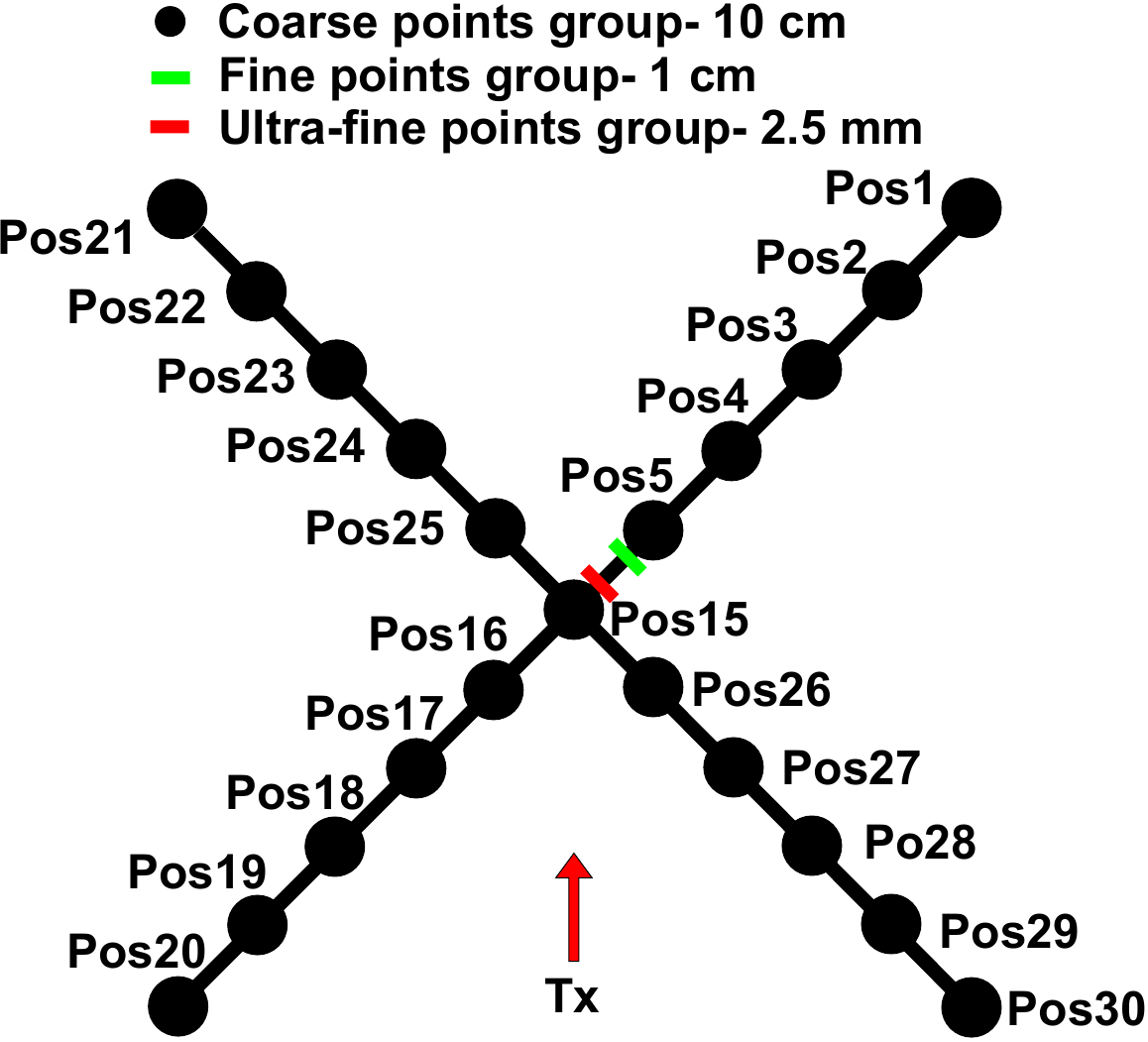}
    \caption{Measurement points along the cross-shaped trajectory. Black circles indicate coarse sampling with 10~cm spacing, while green and red ribbons denote fine and ultra-fine regions with 1~cm (from position~5 toward~15) and 2.5~mm (from position~15 toward~5) spacing, respectively.}
    \label{fig:cross_shaped_trajectory}
\end{figure}

The channel sounding measurements were conducted at Northeastern University’s Ultrabroadband Nanonetworking Laboratory (UNLab), EXP Building, 7th floor, in a room resembling typical classroom ($20\times12\times3$~m$^3$), see Fig.~\ref{fig:Measurement_ChannelSounding}. The Tx was placed 5~m from the wall and 10~m from the front windows, while the Rx was initially placed at the room center, 5~m from the Tx. The Rx was then moved along a cross-shaped trajectory composed of two orthogonal 1-meter lines intersecting at the center, as shown in Fig.~\ref{fig:Measurement_ChannelSounding} and Fig.~\ref{fig:cross_shaped_trajectory}. The spatial sampling resolutions used are coarse (10~cm), fine (1~cm), and ultra-fine (2.5~mm). The room contained several tables and pieces of equipment distributed around its perimeter that resemble a typical classroom. This configuration enabled analysis of variations in MPCs and PDPs across different Rx positions, allowing investigation of whether measurable differences occurred at spatial intervals comparable to the sub-THz wavelength.

Both Tx and Rx antennas were mounted at a height of $1.5$~m. The Tx antenna remained fixed and oriented towards the center of the cross-shaped trajectory for the entire experiment. At each Rx sampling point, the Rx antenna was swept in azimuth and elevation with $10^\circ$ increments, spanning from $-100^\circ$ to $100^\circ$ in azimuth and from $-20^\circ$ to $20^\circ$ in elevation. Each transmission was repeated ten times per Rx position and sweep angle, enabling noise reduction through coherent averaging and improving the robustness of the extracted channel parameters.

The Rx antenna was intentionally not fully swept over the full 360$^\circ$. While weak MPCs from far-wall reflections may exist in principle, their contribution was assumed negligible for the considered link budget and measurement sensitivity, and thus unlikely to affect the dominant channel characteristics analyzed in this work.

The transmitted sounding waveform was generated by mapping an 8191-chips-long m-sequence header (maximum-length sequence) followed by three repetitions of a 4095-chips-long m-sequence onto binary pulse amplitude modulation (2-PAM) symbols. The resulting sequence was pulse-shaped using a square-root-raised-cosine (SRRC) filter with a high roll-off factor to preserve the Nyquist intersymbol-interference-free property and was digitally upconverted to an IF of 5~GHz at a symbol rate of 5~GSym/s. This design yields a waveform with a sharp autocorrelation function, making it well-suited for high-resolution channel sounding.

At the Rx, the observed signal consists of the transmitted sequence as distorted by the propagation channel, including multipath effects. Cross-correlating the received signal with the known transmitted sequence yields the channel impulse response (CIR). The squared magnitude of this cross-correlation yields the PDP, where strong and weak peaks correspond to LoS and MPCs, respectively.

\subsection{Data Processing}

The first step was to process the measured data to obtain well-shaped PDPs, which serve as the basis for extracting key propagation metrics, including RMS delay spread, RMS angular spread, K-factor, and coherence bandwidth for propagation channel evaluation. To this end, the received signal was correlated with a locally generated m-sequence to extract the CIR. The CIR is then coherently averaged over the 10 repetitions to reduce the observable noise floor and suppress spurious detections. 

\subsubsection{Power Delay Profile}

Measurements performed with directional antennas yield directional CIR. By averaging CIRs over repetitions, the directional PDPs are obtained as
\begin{align}
    P(\tau,\Phi,\theta;d) = \big| h_{avg}(\tau,\Phi_{Tx},\theta_{Tx},\Phi_{Rx},\theta_{Rx};d) \big|^2,
\end{align}
where the terms $\Phi_{Tx}$, $\theta_{Tx}$, $\Phi_{Rx}$, and $\theta_{Rx}$ denote the azimuth angles and the elevation angles of the Tx and Rx antennas, respectively. Here, $h_{avg}(.)$ represents the average directional CIR over repetitions and $\tau$ is the delay bin. 

In this work, the delay axis was truncated using a fixed temporal gate of $\tau_n = 85$~ns, corresponding to the maximum excess delay associated with first-order reflections in the considered indoor geometry. This delay gate defines the temporal support of the propagation channel used for subsequent metric extraction.

The max-directional PDP is the strongest received power at each delay bin $\tau$ for a given separation distance $d$, regardless of the specific Tx and Rx pointing directions, obtained by
\begin{align}
    P_{max-dir}(\tau;d) = \underset{\Phi_{Tx}, \Phi_{Rx},\theta_{Tx},\theta_{Rx}}{\max}P(\tau,\Phi_{Tx},\theta_{Tx},\Phi_{Rx},\theta_{Rx};d).
\end{align}

The omnidirectional PDP is obtained by selecting the strongest path at each azimuth angle and averaging over elevation angles, providing a comprehensive channel representation. In this campaign, only the Rx antenna was swept while the Tx antenna remained fixed and aligned with the LoS direction. The omnidirectional PDP is therefore given by\cite{abbasi2022thz}:
\begin{align}
    P_{omni}(\tau;d) = \underset{\Phi_{Tx}, \Phi_{Rx}}{\max} \sum_{i}\sum_{j}P(\tau,\Phi_{Tx},{\theta}^i_{Tx},\Phi_{Rx},{\theta}^j_{Rx};d),
\end{align}
where $i\in{\{1,2,3,4,5\}}$ represents the elevation index for the Rx sweep angles $\{-20^\circ,-10^\circ,0^\circ,10^\circ,20^\circ\}$ and $j = 1$ as the Tx antenna is fixed towards the center of the cross-shaped trajectory. 

\subsubsection{MPC Detection}

The MPCs were extracted from the both omni- and maxi-directional PDPs. The strongest peak was identified as the LoS component, while additional MPCs were detected based on adaptive threshold validation as 
\begin{align}
    Th(\tau) = \epsilon \ [\alpha \ {\tilde{\mu}_{med}}(\tau) + (1-\alpha) \ \mu_{train}(\tau)],
\end{align}
where $\epsilon$ is the offset factor, ${\tilde{\mu}_{med}}(\tau)$ is the local median of the noise floor estimated in a sliding guard–training window around the delay bin $\tau$, $\mu_{train}(\tau)$ is the local mean of the training samples, $N_{train}$, in the noise region (i.e., $10\%$ tail in the late delay bins), and $\alpha$ is the weighting factor $(0 \leq \alpha \leq 1)$ that balances between the median-based and mean-based noise estimates.

This approach ensures robust and consistent MPC detection at all Rx positions, enabling reliable extraction of subsequent channel metrics. The local noise floor was estimated using a sliding training–guard window combined with moving median filtering, and the detection threshold was set to $\epsilon$ dB above this local noise estimate, as summarized in Table \ref{table_detection_param}.

\captionsetup[table]{labelfont={default},labelformat={default},labelsep=newline,name={TABLE},justification=centering}
\begin{table}[!t]
\vspace{2mm}
\renewcommand{\arraystretch}{1}
\setlength{\extrarowheight}{1pt} 
\caption{\textsc{Parameters for detection.}}
\label{table_detection_param}
\centering
\begin{tabular}{|l|l|}
\multicolumn{2}{c}{\textbf{\normalsize }} \\
\hline
\textbf{\normalsize Parameter} & \textbf{\normalsize Value}\\
\hline\hline
$\epsilon$ & $6$~dB\\
\hline
$\alpha$ & $0.4$\\
\hline
$N_{train}$ & $31$~samples \\
\hline
$N_{guard}$ & $15$~samples \\
\hline
$\tau_n$ & $85$~ns\\
\hline
\end{tabular}
\end{table}

\subsubsection{Channel Metrics}

The RMS delay spread, $\sigma_\tau$, quantifies the dispersion of the channel in the temporal domain. It provides a measure of the variability of the mean delay and is calculated as the second central moment of the PDP, as \cite{abbasi2022thz} 
\begin{align}
    \sigma_\tau = \sqrt{\frac{\sum_{i=0}^{N-1}(\tau_i - \hat{\tau} )^2 \ P_i}{\sum_{i=0}^{N-1} P_i}},
\end{align}
where $i=0$ and $N$ denote the indices of the first and last delay bins above the noise threshold, $\tau_i$ represents delay, and $P_i$ is the power received from the path $i$. The mean delay $\hat{\tau}$ is
\begin{align}
    \hat{\tau} = \frac{\sum_{i=0}^{N-1} \tau_i \ P_i}{\sum_{i=0}^{N-1} P_i}.
\end{align}

Similarly, the RMS angular spread ($AS_{RMS}$) characterizes the spatial multipath richness of the propagation channel and is computed as~\cite{abbasi2022thz}:
\begin{align}
    AS_{RMS} = \sqrt{\frac{\sum_{i=0}^{N-1}(\theta_i - \hat{\theta} )^2 \ P_i}{\sum_{i=0}^{N-1} P_i}},
\end{align}
where $\theta_i$ is the AoA of path $i$ and $\hat{\theta}$ is the mean AoA:
\begin{align}
    \hat{\theta} = \frac{\sum_{i=0}^{N-1} \theta_i \ P_i}{\sum_{i=0}^{N-1} P_i}.
\end{align}

The Rician K-factor characterizes the power distribution between LoS and NLoS components, showing concentration of power in the dominant path relative to the remaining MPCs in the channel. This metric provides insight into the channel fading characteristics and is defined as
\begin{align}
    \kappa_1 = \frac{P(\tau_1)}{\sum^{\tau_N}_{\tau_k = \tau_2} P(\tau_k)},
\end{align}
where $\tau_1$ is the LoS delay and $\tau_k$ are the delays of the remaining MPCs ordered by magnitude.

The coherence bandwidth $B_c$ is estimated from the RMS delay spread using the approximation $B_c \approx 1/(5\sigma_\tau)$ (corresponding to a 0.5 correlation threshold). It defines the frequency range over which the channel can be considered flat fading.

\section{Results and Discussion}\label{sect:results}

We begin with the PDP statistics, which highlight the overall behavior of the channel at different measurement locations. We then analyze the channel statistical parameters, followed by an examination of the spatial autocorrelation~\cite{ju_2021_140} statistics among the channel parameters.

\begin{figure}[t!]
    \centering
    \includegraphics[width=1\linewidth]{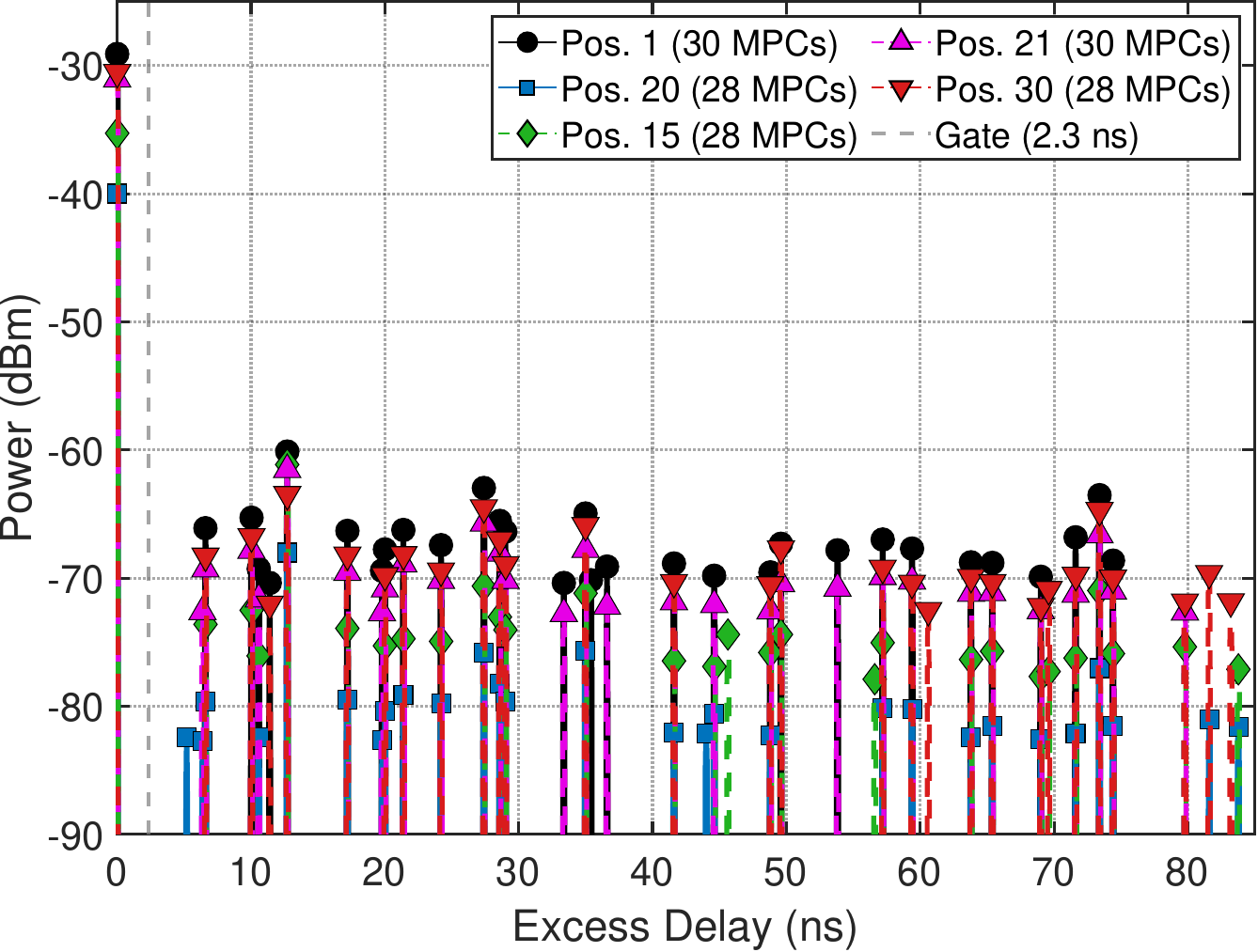}
    \caption{Max-directional PDPs at the center and four endpoints of the trajectory (positions 15, 1, 20, 21, and 30; see Fig.~\ref{fig:cross_shaped_trajectory}), showing the discrete MPCs resolved within the observation window.}
    \label{fig:max-dir_omni-PDPs}
    \vspace{-3mm}
\end{figure}

Fig. \ref{fig:max-dir_omni-PDPs} presents max-directional PDPs at five positions along the measurement trajectory, including the center point and the four endpoints, see Fig. \ref{fig:cross_shaped_trajectory}. The PDPs reveal a highly directional channel dominated by the LoS path. The vertical line at $2.3$~ns marks reflections from nearby setup components such as the table and mountings. This feature appears consistently across the measured PDPs within the same excess delay range, and no other reflecting surfaces were present in proximity to the setup within the corresponding distance that could reasonably account for such reflection.

Channel metrics were derived from the gated and denoised max-directional PDP, except for the RMS angular spread, which was obtained from the directional PDPs. In this work, MPCs originating from reflections off the measurement setup (e.g., the table and mounting structures) were excluded to better capture the intrinsic characteristics of the propagation channel. As a result, channel metrics were computed based on this definition. While the average RMS delay spread across all Rx positions is 6.1~ns when including all detected components, the reported value (22.3~ns) corresponds to the intrinsic channel definition adopted in this work. Similarly, the mean measured K-factor is 16~dB when all components are considered, whereas a value of 18.5~dB is obtained under the same definition. Both metrics consistently indicate a highly directional, LoS-dominated channel.

\begin{figure}[t!]
    \vspace{-5mm}
    \centering
    \includegraphics[width=1\linewidth]{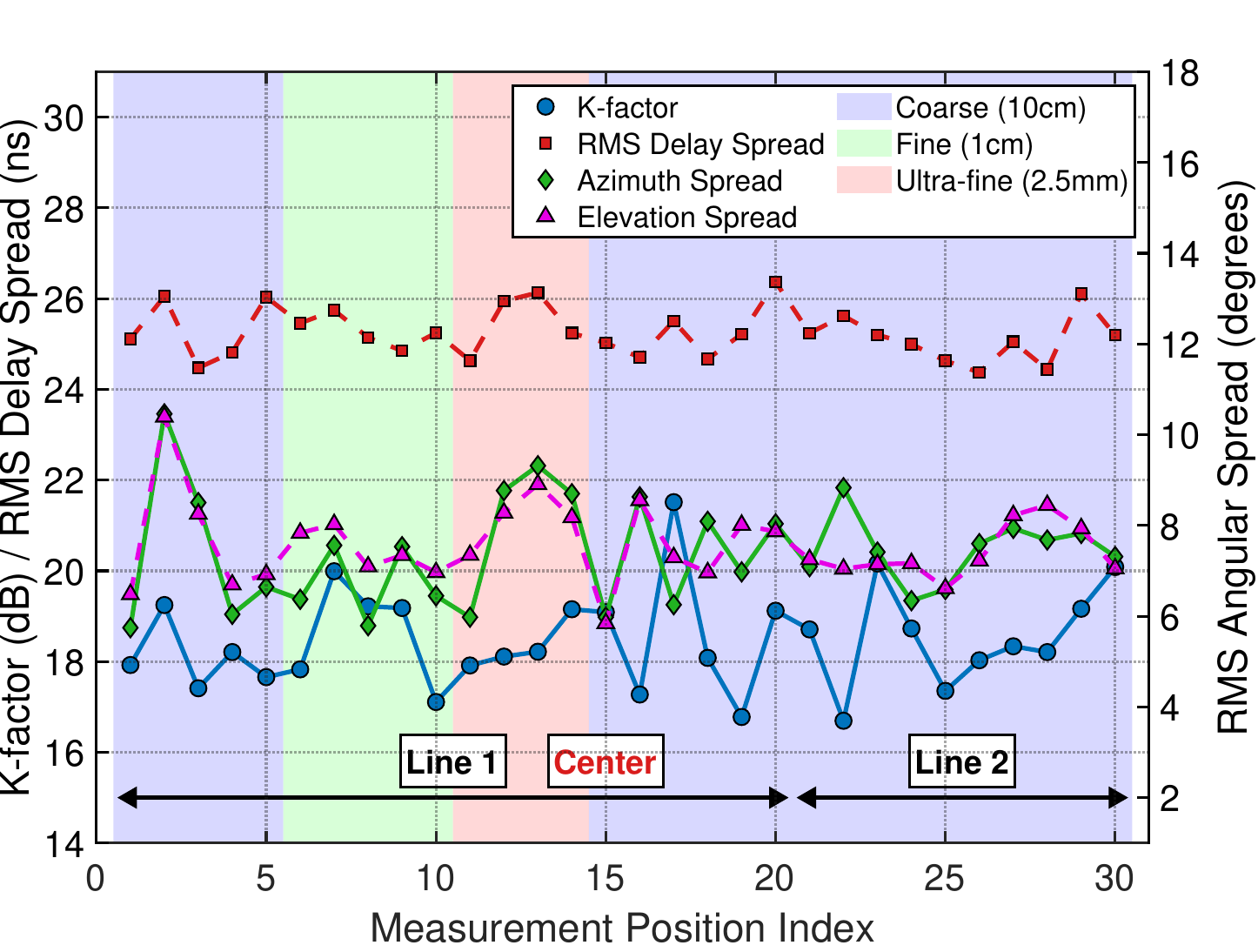}
    \caption{K-factor, RMS delay spread, and RMS angular spreads versus position index along 30 measurement points. Background colors denote coarse (blue), fine (green), and ultra-fine (yellow) spacing intervals.}
    \label{fig:All_metrics_dual_axis}
\end{figure}

Fig. \ref{fig:All_metrics_dual_axis} shows the spatial distribution of channel metrics along the cross-shaped trajectory. Based on the RMS delay spread, the average coherence bandwidth is estimated to be $9$~MHz. These results are consistent with the directional nature of the PDPs, confirming that the channel exhibits strong directivity. The RMS angular spreads further indicate a highly directional channel, with an RMS azimuth spread of $7.4^\circ$ ($\pm 1.2^\circ$) and an elevation spread of $7.6^\circ$ ($ \pm 0.88^\circ$), following the same pattern as the horn antenna. Since the RMS angular spread depends on the power distribution across measurement angles, nearly identical values are expected, as both the Tx and the Rx employ conical horn antennas with comparable half-power beamwidths (HPBWs) in azimuth and elevation. This does not imply that antenna directivity determines the intrinsic RMS angular spread of the propagation channel. Rather, the antenna patterns act as angular filters, shaping the observed angular power distribution and, consequently, the measured RMS angular spread. 

\begin{figure}[t!]
    \vspace{-0mm}
    \centering
    \includegraphics[width=1\linewidth]{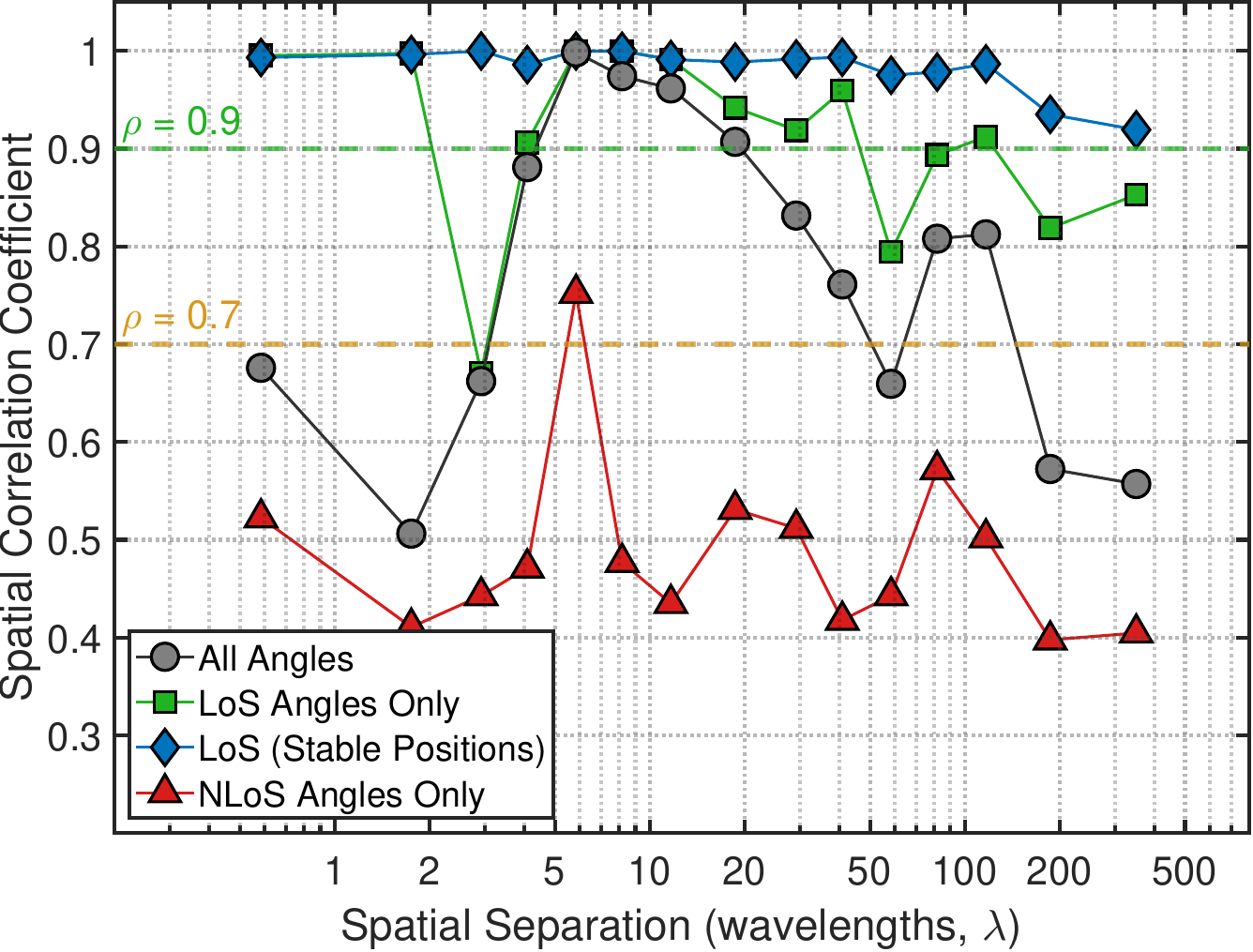}
    \caption{Pairwise directional PDP spatial autocorrelation between measurement points at different distances, categorized into 15 pairwise distance groups.}
    \label{fig:decorrelation}
    \vspace{-0mm}
\end{figure}

To assess spatial consistency between measurement points along the cross-shaped trajectory, we first evaluated the consistency of individual CIRs at each position. The results showed high consistency, with correlation coefficient values $\rho$ exceeding 0.95, confirming that the measurements are stable and not noise-dominated. Next, autocorrelation for different position pairs were analyzed in three categories based on Rx positions, with 22 coarse (10~cm), 6 fine (1~cm), and 5 ultra-fine (2.5~mm) spacing---covering all 435 position pairs and separations from 1.2~$\lambda$ to 483~$\lambda$ (2.5~mm--1~m). Each group captures a distinct scale of channel variation, with the sub-wavelength meshed group providing valuable insight into the de-correlation distance (see Fig.~\ref{fig:decorrelation}). 

Omni-directional PDPs were used to assess overall channel correlation, while directional PDPs evaluated angular consistency. Spatial autocorrelation analysis showed that omni-directional PDPs exhibited near-perfect autocorrelation ($\rho \approx 1$), indicating that averaging across directions masks the spatial and angular dependencies of the sub-THz channel. Spatial autocorrelation values for individual angular PDPs also remained near-perfect. To further examine the origins of spatial de-correlation, directional measurements were decomposed into LoS and NLoS subsets. LoS components---empirically defined as those within 15~dB of the main beam and $\geq 50 \%$ of energy in the first 3~ns---exhibited strong spatial autocorrelation ($\rho > 0.8$) when evaluated at angles qualifying as LoS at both positions, confirming the stability of dominant paths. When restricting the analysis to LoS angles that remained common across paired positions (LoS stable positions), the spatial autocorrelation further increased ($\rho > 0.9$), isolating angularly aligned and stable LoS propagation. In contrast, the all-angles spatial autocorrelation reflects the combined contribution of aligned and misaligned LoS angles together with NLoS components, resulting in consistently lower values than the LoS-only cases despite remaining relatively high due to LoS dominance. NLoS components showed pronounced de-correlation ($\rho \approx 0.45$) due to their sensitivity to sub-wavelength displacements. These findings indicate that overall spatial autocorrelation arises from the interplay between stable LoS and rapidly varying NLoS contributions, suggesting that ray-tracing meshing should adapt to local propagation conditions---using coarser spacing along stable LoS beams and finer resolution in scattered regions---to balance physical fidelity and computational efficiency.

\section{Conclusions}

The aim of this paper was to evaluate the degree of spatial consistency of the sub-THz channels in a complex InH environment. To this end, a detailed measurements campaign was conducted in the 140--150~GHz band covering separations of up to 1~m. As measures, we utilized the spatial autocorrelation coefficient of several key channel characteristics, including PDP, RMS delay spread, RMS angular spreads, and K-factor.

The results show that channel characteristics---including delay, azimuth, and elevation spread, as well as K-factor---exhibit only minor variations across measurement points along the cross-shaped trajectory. Increasing the separation distance between points may lead to larger variations in these metrics. However, consistency tests indicate that, in the considered InH environment and within the examined LoS region, the spatial correlation distance can extend up to approximately 10--50~wavelengths. Accordingly, the minimum mesh grid step size can be chosen to be about 2--10~cm (at 145~GHz) along stable LoS directions, whereas finer spatial resolution may be required in regions dominated by NLoS scattering. For larger mesh grid spacing, advanced spatial interpolation techniques should be used to accurately capture rapidly varying scattered components. This distinction between stable LoS and fluctuating NLoS regions is also essential for accurate beam tracking and beamforming in sub-THz indoor channels. 

Collectively, the abovementioned findings provide quantitative insights that can be utilized for spatial meshing and model parameterization in ray-based and hybrid sub-THz channel models.


\balance
\bibliographystyle{IEEEtran}
\bibliography{reference}

\end{document}